\documentclass[aps,twocolumn,showpacs]{revtex4}
\usepackage{graphicx}
\usepackage{slashed}
\def\bc{\begin{center}}
\def\ec{\end{center}}

\def\beq{\begin{equation}}
\def\eeq{\end{equation}}

\begin{document}

\title{Frequency splitting of intervalley phonons in graphene}
\author{K. Ziegler$^1$ and E. Kogan$^2$}
\address{$^1$  Institut f\"ur Physik, Universit\"at Augsburg, 86135 Augsburg, Germany\\
$^2$ Department of Physics, Bar--Ilan University, Ramat Gan 52900, Israel}

\date{\today}
\begin{abstract}
We study the thermal distribution of intervalley phonons in a graphene sheet.
These phonons have two components with the same frequency. The degeneracy of
the two modes is preserved by weak electron-phonon coupling. A sufficiently strong 
electron-phonon coupling, however, can result in a splitting into an optical and an acoustic phonon 
branch, which creates a fluctuating  gap in the electronic spectrum.
We describe these effects by treating the phonon distribution within a saddle-point approximation.
Fluctuations around the saddle point indicate
a Berezinskii-Kosterlitz-Thouless transition of the acoustic branch. 
This transition might be observable in the polarization of Raman scattered light. 
\end{abstract}

\maketitle


Graphene, a two-dimensional sheet of carbon atoms forming a honeycomb lattice,
has highly unusual electronic properties \cite{Novoselov2005,zhang05,Novoselov2006,Katsnelson2006}.
This is due to the fact that there are two bands that touch each other at two Dirac nodes.
Moreover, the low-energy
quasiparticles of undoped graphene experience a linear dispersion around the
Dirac nodes. Transport properties, characterized by the longitudinal conductivity at the Dirac nodes,
are quite robust and do not vary much from sample to sample. Exactly at
the Dirac point a minimal DC conductivity has been observed in a number of experiments
\cite{Novoselov2005,zhang05,Novoselov2006}. Its value is not much affected by disorder \cite{chen08,morozov08}.
A typical source of  disorder in graphene are frozen lattice deformations (ripples) \cite{C,abergel10}. In a more realistic
description, these deformations
may not be frozen but fluctuate thermally due to the softness of the two-dimensional material.

In this paper we investigate effects of the electron--phonon interaction in graphene.
Considering only in-plane displacements of graphene atoms, we have four different phonon modes.
For low-energy electronic states
electron--phonon interaction is efficient if the phonon wavevector is close to $\Gamma$, $K$ or $K'$ points.
As it was explained by Basko and Aleiner \cite{basko08}, the only modes which are effectively coupled 
to electrons are the pseudovector optical phonons corresponding to the irreducible representation $E_2$ of the 
group $C_{6v}$ from the $\Gamma$ point, and the scalar  phonons corresponding to the irreducible representations 
$A_1$ and $B_1$ of the same group from the points $K$ and $K'$.

Experimental information on
the interaction is obtained by Raman spectroscopy \cite{ferrari06,gupta07,graf07,pisana07,yan07} and by 
angle--resolved photoemission 
spectroscopy \cite{bostwick07}. The above mentioned theoretical analysis of the role of different phonons is confirmed by 
the fact that in the Raman spectrum of graphene only two two--phonon peaks are seen:  the  $D^*$ peak  and the  $G^*$ 
peak, corresponding to scalar and pseudovector phonons, respectively.
In our paper we will consider interaction of electrons only with the scalar phonons.


The physics of the electron-phonon system is defined by the Hamiltonian with optical (Einstein)
phonons at frequency $\omega_0$ \cite{sasaki08}:
\beq
H= \omega_0 \sum_{\bf r} b_{\bf r}^\dagger b_{\bf r} 
+ \sum_{{\bf r},{\bf r}'}[h_{{\bf r},{\bf r}'} 
+\alpha(b_{{\bf r},{\bf r}'}+b^\dagger_{{\bf r},{\bf r}'})] c^\dagger_{\bf r} c_{{\bf r}'}
\ .
\label{hamilton00}
\eeq
Here $c^\dagger_{\bf r}$ ($c_{\bf r}$) are the electron creation (annihilation) operators and 
$b^\dagger_{{\bf r},{\bf r}'}$ ($b_{{\bf r},{\bf r}'}$) are the phonon creation (annihilation) operators.
This Hamiltonian describes an effective
attraction between the fermions and a renormalization of the electronic hopping amplitude $h_{{\bf r},{\bf r}'}$, 
reducing the hopping rate substantially (polaron effect). The attractive interaction may 
lead to the formation of Cooper pairs. In 2D this effect is not relevant due to strong 
fluctuations, preventing the system to become superconducting. 

The conventional approach to determine the properties of the phonons and the electrons is based on
a self-consistent evaluation of the self energy (Migdal approximation) 
\cite{park07,deveraux07}. The latter provides an effective (or renormalized)
energy and its imaginary part an effective scattering rate. Such a static approximation might be 
insufficient in a two-dimensional system, since it does not take into account thermal fluctuations. 
This was already discussed in an experimental study of graphene \cite{pisana07}. 
To avoid this problem, we include thermal fluctuations in our approach.
To this end, we replace the phonon operators $b_{{\bf r},{\bf r}'}$, $b^\dagger_{{\bf r},{\bf r}'}$ by their
quantum average: $b_{{\bf r},{\bf r}'}\approx \langle b_{{\bf r},{\bf r}'}\rangle\equiv v_{{\bf r},{\bf r}'}$ and 
$b_{{\bf r},{\bf r}'}^\dagger\approx \langle b^\dagger_{{\bf r},{\bf r}'}\rangle\equiv v_{{\bf r},{\bf r}'}^*$. 
In this approximation we can keep thermal fluctuations but ignore quantum fluctuations of the phonons.
The electrons, on the other hand, are studied in full quantum dynamics. This reduces the
grand-canonical ensemble at inverse temperature $\beta$, defined by the generating function
$Tr e^{-\beta H}$ to a functional integral with respect to thermal fluctuations of the lattice distortations
$u_{{\bf r},{\bf r}'}$ and a trace with respect to the electrons. It should be noticed that only the real part
$Re(u)$ couples to the electrons. After performing the trace over the electrons
we get
\beq
Tr e^{-\beta H}
\approx\int \det({\bf 1}+e^{-\beta h})e^{-\beta S_0}   
{\cal D}[{\vec u}]\equiv Z
\ .
\label{genfunc1}
\eeq
with the dispersion for the rescaled phonon field $u_{{\bf r},{\bf r}'}=\alpha v_{{\bf r},{\bf r}'}$ 
\[
S_0=\frac{\omega_0}{2\alpha^2}\sum_{{\bf r},{\bf r}'}u_{{\bf r},{\bf r}'}^2 
\ .
\]

It is convenient to introduce a sublattice representation for the tight-binding Hamiltonian $h$
and to expand it around the two valleys $K$ and $K'$: 
The graphene unit cell contains two atoms, each of them has one $\pi$-orbital.
This gives a two-component wavefunction. Moreover, the bandstructure has two nodes 
(or valleys). Expansion around the valleys leads to a wavefunction that is 
represented by a 4-component column vector. 
We shall work in a direct product vector space of the valley space 
and the sublattice space. For the phonons we consider here only the intervalley 
contribution, which play a major role in the electron-phonon interaction \cite{basko08}. 
This gives for the low-energy Hamiltonian
\beq
h=-it \Pi_3\slashed{\partial}+{\vec u}\cdot{\vec \Pi}\Sigma_0
\ .
\label{hamiltonian0}
\eeq
The parameter $t$ is the bandwidth. Here we have used a coordinate system that refers to one sublattice and
one valley in the notation of Ref. \cite{mcann}:
the indices of the Pauli matrices $\Sigma_j$ ($j=0,...,3$) are acting on the A-B sublattices and the indices of the Pauli
matrices ($\Pi_j$ ($j=0,...,3$) are acting on the two valleys $K$ and $K'$). 
Moreover, we have $\slashed{\partial}={\vec\Sigma}\cdot{\vec \partial}$, where ${\vec \partial}$ is a lattice difference operator.
For graphene in a homogeneous magnetic field this electron-phonon interaction was considered 
in Ref. \cite{nomura09}. 
Then the intervalley phonon field has two component:
\[
\pmatrix{
0 & u_{\bf r} \cr
u'_{\bf r} & 0 \cr
}
=u_{1,{\bf r}}\Pi_1+u_{2,{\bf r}}\Pi_2 
\ ,
\]
where $u_{1,{\bf r}}=(u_{\bf r} +u'_{\bf r})/2$ and  $u_{1,{\bf r}}=-i(u_{\bf r} -u'_{\bf r})/2$ \cite{sasaki08}.

The integral $Z$ in Eq. (\ref{genfunc1})
serves as a generating function that allows us to get, for instance, the static electronic
Green's by differentiation of $\ln Z$ as
\beq
G_{{\bf r},{\bf r}'}=\frac{1}{Z}\int ({\bf 1}+e^{-\beta h})^{-1}_{{\bf r},{\bf r}'} e^{-\beta S}{\cal D}[{\vec u}]
\label{fint}
\eeq
with 
\beq
S=S_0-\beta^{-1}\log\det({\bf 1}+e^{-\beta h}) , \ \  {\vec u}=(u_1,u_2)
\ .
\label{action0}
\eeq
It is useful to notice that $e^{-\beta S}=e^{-\beta S_0}\det({\bf 1}+e^{-\beta h})$ 
is a non-negative function. Therefore, $e^{-\beta S}/Z$ is a probability
density for the phonon field, and the static one-particle Green's function then 
can also be written as an average $\langle ... \rangle_{ph}$ with respect to 
the distribution $e^{-\beta S}/Z$ \cite{ziegler05}:
\beq
G_{{\bf r},{\bf r}'}=\langle ({\bf 1}+e^{-\beta h})^{-1}_{{\bf r},{\bf r}'} \rangle_{ph}
\ .
\label{average_green}
\eeq


The   
distribution $e^{-\beta S}/Z$ is invariant under a unitary transformation.
For the special transformation
\beq
h\to U h U^\dagger , \ \ \ U=\pmatrix{
1 & 0 \cr
0 & e^{-i\varphi} \cr
}\Sigma_0
\label{usymmetry}
\eeq
with $0\le \varphi < 2\pi$ the electron-phonon coupling term in Eq. (\ref{hamiltonian0})
satisfies the relation
\beq
U{\vec u}\cdot{\vec \Pi}U^\dagger=O{\vec u}\cdot{\vec \Pi}
\ ,
\label{orthogonal}
\eeq
where $O$ is the orthogonal transformation (i.e. rotation by angle $\varphi$) in the space of 
the matrices $\Pi$:
\[
O=\cos \varphi\Pi_0\Sigma_0+i\sin \varphi \Pi_2\Sigma_0
=\exp(i\varphi\Pi_2)\Sigma_0
\ .
\]
Since the gradient part of the Hamiltonian $h$ is invariant under a global 
rotation of the phonon field ${\vec u}$, the distribution  $e^{-\beta S}/Z$ is also invariant.
As a consequence, the phonon field can produce massless fluctuations.


Second order perturbation theory is the standard approach for evaluating the change of the 
phonon frequency by the electron-phonon interaction \cite{C,castro07,stauber08}.
In our model, defined by the action (\ref{action0}), second order in $\alpha$ gives us for 
the renormalized frequency of the intervalley mode
\beq
\alpha^2\frac{\partial^2 S}{\partial u_i\partial u_j}\Big|_{u=0} 
=(\omega_0-a_2)\delta_{ij}
\label{pert_theory}
\eeq
with
\beq
a_2=\frac{\alpha^2}{\pi\beta t^2}\ln\left[\frac{\cosh(\beta E_F)+\cosh(\beta t\Lambda)}
{1+\cosh(\beta E_F)}\right]
\ ,
\label{coefficient}
\eeq
where $\Lambda=2\sqrt{\pi}$ is the momentum cutoff.
For $\beta t\gg 1$, which is satisfied even at room temperature due to $t\approx 2.7$ eV, 
the renormalized phonon frequency has the asymptotic behavior
\[
\omega_u\sim \omega_0 -\frac{\alpha^2}{t^2\pi}\left(2\sqrt{\pi}t-|E_F|)\right)
\ .
\]
Thus the intervalley phonons are softened by the electron-phonon coupling. Moreover, the phonons hardens
as we go away from the Dirac point $E_F=0$, in agreement with recent experiments 
\cite{yan07,pisana07,stampfer07}.
However, this frequency becomes negative if $\omega_0<a_2$, indicating an instability
of the electron-phonon system. In particular, the vanishing frequency reveals a phonon softening, where the
optical phonon mode becomes acoustic.
In the following we shall discuss that this instability is associated with a 
splitting of the degenerate phonon modes, where instead of a single phonon frequency two different 
frequencies appear.


The instability cannot be treated within second-order perturbation theory but requires
a self-consistent approach. Here it is natural to perform the integration in Eqs. (\ref{genfunc1})
and (\ref{fint}) in a saddle-point (SP) approximation. This leads to the SP equation $\delta_uS=0$
which determines an average lattice distortion $|{\vec u}|={\bar u}$.
 Assuming a uniform solution ${\bar u}$ the SP equation $\partial S/\partial u_i=0$  reads
\beq
{\bar u}=2\frac{\alpha^2}{\omega_0}{\bar u}\int\frac{1}{\epsilon_k}
\frac{\sinh(\beta\epsilon_k)}{\cosh(\beta E_F)+\cosh(\beta \epsilon_k)}
\frac{d^2k}{(2\pi)^2} 
\label{spa1}
\eeq
with the dispersion $\epsilon_k=\sqrt{t^2k^2+{\bar u}^2}$ of the electrons. A nonzero ${\bar u}$
opens an electronic gap $\Delta=2{\bar u}$. 
This equation fixes only ${\bar u}$, not the direction of the
vector ${\vec u}_0\equiv ({\bar u}\cos\varphi,{\bar u}\sin\varphi)$. For a non-trivial solution ${\bar u}\ne0$ we 
we can perform the integration in Eq. (\ref{spa1}) and obtain the equation
\beq
\frac{\cosh(\beta E_F)+\cosh(\beta\sqrt{\Lambda^2t^2+{\bar u}^2})}{\cosh(\beta E_F)+\cosh(\beta {\bar u})}
=e^{\beta\pi t^2\omega_0/\alpha^2}
\ .
\label{spe1}
\eeq
Then the critical point $\beta_c$ of the instability is determined from this equation for ${\bar u}=0$.


For $\beta t\gg 1$ and $|E_F|<{\bar u}$ we get from Eq. (\ref{spe1})
\beq
{\bar u}\sim\frac{2\alpha^2}{\omega_0}\left(1-\frac{\pi t^2\omega_0^2}{4\alpha^4}\right)
\label{order_p1}
\eeq
and for $\beta t\gg 1$, $|E_F|>{\bar u}$
\beq
{\bar u}\sim \sqrt{\left(\frac{\pi t^2 \omega_0}{\alpha^2}+|E_F|\right)^2-4\pi t^2}
\ .
\label{order_p2}
\eeq
While an increase of the electron-phonon coupling $\alpha$ increases ${\bar u}$ for 
$|E_F|<{\bar u}$, it decreases ${\bar u}$ for $|E_F|>{\bar u}$. This result indicates that
the instability of the ${\bar u}=0$ solution is supported by the electron-phonon interaction
near the Dirac point, whereas the instability is suppressed further away from the Dirac point.

\begin{figure}[t]
\begin{center}
\includegraphics[width=8cm,height=6cm]{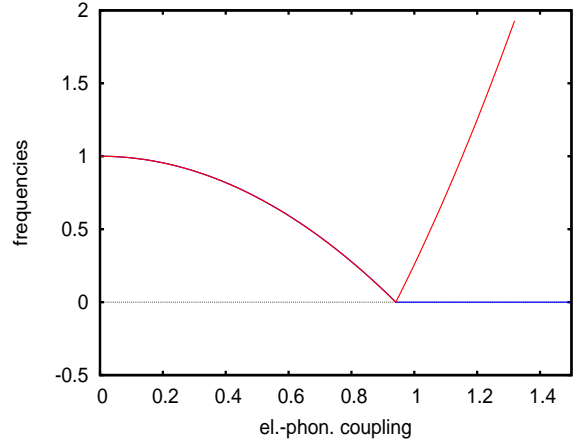}
\caption{Splitting of the phonon modes (from Eqs. (\ref{ren_frequency1}, (\ref{ren_frequency2})).
The red (blue) curve represents $\omega_1$ ($\omega_2$) in units of $\omega_0$, indicating that 
for the dimensionless electron-phonon coupling $\alpha/\sqrt{t\omega_0}>a_c\approx 0.94$ there exist 
two different frequencies.}
\label{fig:splitting}
\end{center}
\end{figure}

For small ${\bar u}$ Eq. (\ref{spe1}) can be expanded. This allows us to study the properties
of the phonon renormalization for nonzero ${\bar u}$, i.e. in the regime where ${\bar u}=0$ and
the perturbation theory are unstable. The SP equation then reads
\beq
\alpha^2\frac{\partial S}{\partial u_j}\Big|_{u={\bar u}} 
\approx (\omega-a_2 + a_4 {\bar u}^2){\bar u}_j =0
\ .
\label{spe2}
\eeq
Besides the solution ${\bar u}=0$ this equation has the nonzero solutions
\[
{\bar u}=\pm\sqrt{a_2-\omega_0} /\sqrt{a_4} \ \ \ (a_2-\omega_0\ge 0)
\ .
\]
with
\beq
\alpha^2\frac{\partial^2 S}{\partial u_i\partial u_j}\Big|_{u={\bar u}}
=4(a_2-\omega_0)\frac{{\bar u}_i {\bar u}_j}{{\bar u}^2}
\label{curve2}
\eeq
instead of (\ref{pert_theory}). 
Since the SP solution ${\vec u}_0$ is on a circle with radius ${\bar u}$, the eigenvalues
of the matrix on the right-hand side are $\omega_1=4(a_2-\omega_0)$ and $\omega_2=0$. 
For a stable solution, on the other hand, ${\bar u}=0$ the frequency shift agrees
with the result of the second-order perturbation theory.
Combining (\ref{pert_theory}) (for $a_2<\omega_0$) and 
(\ref{curve2}) (for $a_2\ge \omega_0$), we obtain the renormalized phonon frequencies as
\beq
\omega_1  
\approx |a_2-\omega_0|\left[1+\Theta(a_2-\omega_0)\right]
\label{ren_frequency1}
\ ,
\eeq
\beq
\omega_2\approx (\omega_0-a_2)\Theta(\omega_0-a_2)
\  .
\label{ren_frequency2}
\eeq
Thus, the mode with $\omega_2$ vanishes for the solution ${\bar u}\ne 0$, as plotted in
Fig. \ref{fig:splitting}.

After having solved the SP equation for a uniform phonon field ${\vec u}$,  
we study the fluctuations around the uniform solution. Fluctuations may play an important role and should
modify the uniform solution due to the two-dimensionality of graphene. In particular, the fluctuations
are crucial for the vanishing phonon frequency $\omega_2$ in Eq. (\ref{ren_frequency2}).

With the special SP solution ${\vec u}_0$ also any rotated phonon mode $O{\vec u}_0$ is a SP solution. 
Therefore, we can choose the special solution ${\vec u}_0=({\bar u},0)$
and, using the identity (\ref{orthogonal}), obtain the action at the SP as
\[
S\approx S_0-\frac{1}{\beta}\log\left[\det({\bf 1}+e^{-\beta(-it\Pi_3\slashed{\partial}+{\bar u}U\Pi_1U^\dagger)})
\right]
\ ,
\]
which is degenerated with respect to a global unitary transformation $U$. 
In order to study fluctuations around the SP solution we
introduce a spatially fluctuating unitary matrix $U_r$.
This can be cast into a nonlinear matrix field
\beq
\Phi_{\bf r}= U_{\bf r}\Pi_1U_{\bf r}^\dagger=\pmatrix{
0 & e^{i\varphi_{\bf r}} \cr
e^{-i\varphi_{\bf r}} & 0 \cr
}\Sigma_0
\ ,
\label{matrixfield}
\eeq
such that the action becomes 
\beq
S\approx S_0-\frac{1}{\beta}\log\left[\det({\bf 1}+e^{-\beta(-it\Pi_3\slashed{\partial}+{\bar u}\Phi)})\right]
\ .
\label{effaction}
\eeq
This action can be expanded, either for ${\bar u}/t< 1$ or for ${\bar u}/t> 1$.

Now we assume that ${\bar u}>t$. Formally, $t$ can be chosen independently of the real 
bandwidth of the system.
But this choice means that the physics is restricted to quasiparticles up
to energy $t$. Then the expansion of the action $S$ of Eq. (\ref{effaction})
in powers of $1/{\bar u}$ 
provides a $U(1)$ nonlinear sigma model:
\beq
S\approx S_0'   
+F(\beta, {\bar u})
Tr(\slashed{\partial}\Phi\slashed{\partial}\Phi)
\label{action2}
\eeq
with the prefactor 
\beq
F(\beta ,{\bar u})
=\frac{t^2}{{\bar u}^2}\left[
\frac{{\bar u}}{4}+\frac{1}{\beta}\ln(1+e^{-\beta {\bar u}})+\frac{1}{2}\frac{{\bar u}}{1+e^{\beta {\bar u}}}
\right]
\ .
\label{prefactor}
\eeq
Using the result of ${\bar u}$ for $\beta {\bar u}>\beta t\gg 1$, 
the $\beta$ dependence drops out of the prefactor: 
\beq
F\sim\frac{t^2}{4{\bar u}}
\sim\frac{t^2\omega_0}{32\pi \alpha^2}\frac{1}{1-t^2\omega_0^2/16\alpha^4}
\ .
\label{prefactor1}
\eeq
This result can be considered as a renormalization effect for the temperature, where
the renormalized dimensionless temperature reads
\beq
\tau=\frac{\bar u}{t^2}T
\sim\frac{2\alpha^2}{\omega_0 t^2}\left(1-\frac{\pi t^2\omega_0^2}{4\alpha^4}\right) T
\ .
\label{ren_temp1}
\eeq
Here we have assumed $|E_F|<{\bar u}$ and have used the expression of ${\bar u}$ in
Eq. (\ref{order_p1}).
As the electron-phonon coupling $\alpha$ increases, the renormalized temperature $\tau$
increases as well. This reflects the fact that the electronic fluctuations
enhance the phonon fluctuations. The case $|E_F|>{\bar u}$ is rather unrealistic here, 
since we also have assumed ${\bar u}>t$.

The fluctuating term in Eq. (\ref{action2}) represents an XY model for the angular fluctuations
$\varphi_r$. Thus, the phonon fluctuations undergo a Berezinskii-Kosterlitz-Thouless (BKT) transition
if $\tau_c$ is of order one 
\cite{bkt}. Thus, for temperatures below the BKT transition point $\tau_c$ the fluctuations are strongly
correlated whereas above this temperature the correlations of the fluctuations decay exponentially
due to the proliferation of vortex pairs.


We conclude from our calculation that
second-order perturbation theory with respect to electron-phonon interaction reveals
a phonon softening. This result is in agreement with other calculations \cite{C,castro07,stauber08}.
For sufficiently large electron-phonon coupling the perturbative approach breaks down and a self-consistent
calculation is necessary. We have used an SP approximation for the phonon fluctuations and found
an instability of the perturbative approach due to a splitting of the optical phonon mode into an optical
and an acoustic branch. The latter is related to a massless mode caused by the rotational symmetry of the 
two system. It is characterized by long-ranged correlations of the phonon fluctuations, in contrast to the
short-range fluctuations of the optical phonons. However, at sufficiently high temperatures these long-range
correlations can undergo a BKT transition by the creation of vortex pairs, resulting
again in short-range correlated fluctuations. The average lattice distortion vanishes for all regimes as
a consequence of the Mermin-Wagner argument.
A possible way to observe the BKT transition experimentally is to measure the polarization in 
Raman scattering, since the photon polarization couples to the direction of the lattice distortions. 

The instability of the intervalley phonons is very different from the instability of the Holstein
(out-of-plane) phonons studied recently, where the phonons undergo an Ising transition \cite{ziegler10b}.
In the latter case only short-range correlated fluctuations appear (except for the critical point) but the average
distortion has  a non-zero value. 

To understand the effect of the thermal phonon
fluctuations on the electronic transport properties, we can return to Eq. (\ref{average_green}). 
Then ${\vec u}_{\bf r}$ produces a random gap in the electronic spectrum which has to be averaged
with respect to the phonon distribution $e^{-\beta S}/Z$.
Thus, the effect of thermal phonons is similar to the effect of frozen 
correlated disorder, depending on the temperature of the sample though. 
Since an uncorrelated random gap with vanishing mean does not affect the minimal
conductivity \cite{ziegler09,medvedyeva10,ziegler10a}, we expect a similar result for 
thermally fluctuating intervalley phonons, at least near the Dirac point.

In conclusion, our calculation reveals that the electron-phonon interaction in graphene 
leads to a substantial renormalization of the intervalley phonons. For sufficiently strong
electron-phonon coupling this causes an instability, where the phonon frequency vanishes and
a new pair of phonons appears, consisting of an optical and an acoustic branch.

\section*{ACKNOWLEGEMENTS}

We acknowledge financial support by the DFG grant ZI 305/5-1.

\end{document}